\documentclass[sigconf]{acmart}

\usepackage[]{algorithm2e}

\usepackage{amsmath}
\usepackage[skip=10pt]{caption}

\usepackage{balance}

\begin{document}

\title{Local versus Global Strategies in Social Query Expansion}

 \author{Omar Alonso}
  \affiliation{Microsoft} 
 \email{omalonso@microsoft.com}
 \author {Vasileios Kandylas}
\affiliation{Microsoft}
\email{vakandyl@microsoft.com}
 \author{Serge-Eric Tremblay}
  \affiliation{Microsoft} 
 \email{sergetr@microsoft.com}
 
\begin{abstract}
Link sharing in social media can be seen as a collaboratively retrieved set of documents for a query or topic expressed by a hashtag. Temporal information plays an important role for identifying the correct context for which such annotations are valid for retrieval purposes. We investigate how social data as temporal context can be used for query expansion and compare global versus local strategies for computing such contextual information for a set of hashtags.
\end{abstract}

\maketitle

\section{Introduction}

Twitter and Facebook are the largest sources of public opinion and real time information about any topic anywhere on the world.  Trending topics and hashtags provide a very strong signal for popular events that  later would be covered in news articles while, at the same time, controversial links tend to trigger more discussion and polarization on social networks. 

We use human sensing on Twitter as a large distributed human crawler where links are constantly shared and annotated. 
Users' comments or social annotations are human generated content that can not only provide  additional context but also offer 
snapshots in time by capturing the   social chatter vocabulary.
An example
of a common annotation is the inclusion of a hashtag on a post to emphasize or tag the content with an entity or event. 
This large scale link selection can be seen as a highly curated human retrieved set of documents for a query as expressed by a hashtag. 

Automatic query expansion is a well known method for improving information retrieval systems where the user's original query is augmented by new features with similar meaning. 
 By capturing a temporal context from Twitter, we can extend techniques like pseudo relevance feedback and query expansion to include social signals for retrieval purposes.
The driving scenario for our work is the extraction of relevant links from Twitter that cover the core of an event (described by one or more hashtags) as it evolves over time by using selected relevant contend derived from social data. 

In this paper we investigate how social data as  temporal context can be used for query expansion and 
the difference between  global versus local methods of expansion. Little work has been done investigating the use of tweets alone for automatic query expansion when retrieving shared links or documents  from microblogs.
There is previous work on using external sources for query expansion  \cite{WangHF17, ZinglaLMBS18} and
anchor text~\cite{DangC10} in the context of Twitter.
In contrast, we rely on Twitter data only for term extraction and weights without any manual intervention nor external sources like knowledge bases or Wikipedia. 
 We also use anchor text, but  derived as a form of social anchor text for extracting candidate terms. 
Related to our work, a two-step extension for computing pseudo-relevance feedback  that contains a manual tweet selection and query-document temporal relevance model is described in \cite{MiyanishiSU13}.
There are similarities in our approach to local context analysis, a blind relevance feedback technique described in~\cite{Xu00}.

\section{Methods}

We briefly describe the main methods that we employed in this study that use the Twitter firehose as the only input data source.
 Aggregations on the Twitter data are computed using social signatures and contextual vectors using a fine-grained voting scheme. This allow us to build a hashtag  index that is then used for retrieval purposes where documents are re-ranked using social pseudo relevance feedback (SPRF)  and social  query expansion (SQE) techniques~\cite{AlonsoKT18}.

\subsection{Contextual Vectors and Social Signatures}

A contextual vector represents a ranked list of ngrams (sizes 1 to 4) for a set of tweets related to a hashtag or entity. 
For producing the list of ngrams, we first aggregate all the tweets related to a particular hashtag over the time period of consideration. 
A social signature is a high level representation of a web page from a social media perspective, that is, a list of ngrams associated with the link.
The computation of the contextual vectors is similar to that of the social signatures. The main difference is that social signatures are computed for links instead of hashtags or entities. 
Examples of a contextual vector \texttt{\small{\#charliehebdo =  (free speech, charlie hebdo, terrorist attack, satirical magazine, sad day, paris attack}}) and  social signature \texttt{\small{\url{politico.com/story/2016/12/michigan-attorney-general-files-lawsuit-to-halt-recount-232114} = (fraud total waste, halt recount, Michigan attorney)}}.

Both contextual vectors and social signatures are very sensitive to time. Table~\ref{tab:cv} shows examples of contextual vectors for two hashtags:  \texttt{\small{\#inauguration}} and \texttt{\small{\#london}} in 2017. The inauguration of the President of the US  is a single event with a very descriptive hashtag where the  contextual vectors differ for the three consecutive days but still on the same topic. Comparisons to the previous inauguration ceremony are the subtopics of the main topic. In the case of \texttt{\small{\#london}}, the hashtag is used to describe three very different events that occurred in London at different times: a  terrorist attack in Westminster, another terrorist attack on London Bridge, and a fire at Grenfell Tower.

\begin{table}[h!]
  \centering
  \small {
  \begin{tabular}{| p{1.43cm} | p{0.85cm} | p{5cm}  | }
  \hline
	\textbf{Hashtag}  & \textbf{Date} & \textbf{Contextual vector}      \\
	\hline 
\#inauguration & Jan-19 & marylanders preparing, trump people, jon voights, donald trump, lincoln memorial \\ \hline
\#inauguration &Jan-20 & donald trump, crowds 2009, 2009 inauguration,  2017 inauguration, president trump, 45th president \\ \hline
\#inauguration & Jan-21 & rally tonight, obama 2009, sean spicer, anti trump, trump rally, 250000 obama \\ \hline
\#london & Mar-22 & abu izzadeen, terror attack, westminster bridge  \\ \hline
\#london & Jun-03 & london bridge, terrorist attack, stay strong \\ \hline
\#london & Jun-14 & london fire, grenfell tower, 21st floor\\ 
    \hline
  \end{tabular}
  \caption{Examples of contextual vectors for \texttt{\small{\#inauguration}} and \texttt{\small{\#london}} for 2017.}
  \label{tab:cv}
  }
\end{table}

\subsection{Voting Scheme}

We compute aggregations of social data using frequencies of occurrences of hashtag and their occurrences in tweets and re-tweets.
The problem with using raw frequencies is that spam and advertising accounts  tend to post multiple tweets per day that contain the same information. These behaviors can make links and hashtags  artificially popular, when in fact most of their popularity comes from a small number of Twitter accounts.  We therefore assign a single vote to each account per time period, so that one account cannot skew the frequency of an element or connection.  
For each hashtag,  we compute the following  counts: \texttt{\small{hashtagTweetFrequency}}, \texttt{\small{hashtagRetweetFrequency}}, \texttt{\small{hashtagTotalFrequency}}, \texttt{\small{hashtagTweetVotes}}, \texttt{\small{hashtagRetweetVotes}}, \texttt{\small{hashtagTotalVotes}}. With this fine granular voting scheme we 
capture behavioral data  in a way that can be later use for term weighting. 
Similar counts  are computed for links and ngrams.

\subsection{Hashtag Index}

We produce a hashtag index, a data structure  that contains timestamped information about hashtags, contextual vectors, and social signatures.  The hashtag index gives relevant links
and other information about every selected hashtag and provides a
connection between a hashtag and all the links associated with it. We can think of the hashtag index as a collection of temporal contexts for topics  and links where the indexing key is date (ie., year-month-day) and spans a full year.
The hashtag index also contains similar hashtags (e.g., \texttt{\small{StarWars}} and \texttt{\small{RogueOne}}) that are computed using a SimHash-based algorithm.

\subsection{Social Pseudo Relevance Feedback}

We assume a vector space model where the similarity $sim(q,d)$ between query $q$ and document link $d$ is computed as $sim(q, d) = \sum_{t \in {q \cap d}}{ w_{t,q} . w_{t,d}}$, where $w_{t,q}$ and $w_{t,d}$ are the weights of term $t$ in query $q$ and document link $d$ respectively according to user engagement via RT, likes,
or shares. 
The already mentioned contextual vectors are derived from tweets  and we can think of those ngrams as explicit  terms selected by users in aggregate for a particular time period. Similar to pseudo relevance feedback, ranking documents  extracted from tweets by our voting system indicate that the top-k links are relevant to a given query (i.e., hashtag) on a specific time period, usually daily.

We describe the weighting scheme that is computed for all elements in a contextual vector and social signature. 
Each hashtag, \textit{ht}, has an associated vote record that allows fine tuning to produce the final weight, denoted as counter $c$, for a
hashtag $ht_{c} =
  (\log(\text{ht.tTweetVotes} * p_{w}) + \log(\text{ht.tRetweetVotes} * rt_{w})) * v_{w} + (\log(\text{ht.ltTweetVotes} * p_{w}) +  \log(\text{ht.ltRetweetVotes} * rt_{w})) * l_{w}$, where tweet weight  $p_{w} = 0.8 $, re-tweet weight $rt_{w} = 0.2$, vote weight $v_{w} = 0.35$, and link weight $ l_{w} = 0.5$. 
  That is,
an original term in a tweet has more weight than a re-tweet and a document link has more weight than a regular vote.
Compared to \cite{Xu00}, we rely on a contextual vector for candidate expansion terms instead of extracting them from a search list.
In SPRF we use  social signals for selecting and ranking document that are shared via links in Twitter. Similarly, we can also use social signals for selecting terms for query expansion. 

\subsection{Social Query Expansion}

We use contextual vectors and social signatures to re-rank the links based on how similar the document links' titles are to the terms in 
the contextual vector. Because all these links have been tagged by users with a hashtag with a high vote, our hypothesis is that  
they belong to the same topic. We expand the title matching by also including related hashtags and ngrams from the contextual vectors,  $\text{htTitleScore}  = \text{sim}(\text{title}, \text{q} + \text{ngrams}) $.
A similar formula is applied with respect to RTs.  We then take the $\max$ score from title, description, and file name on the element's dimension (e.g., hashtag, ngram, etc.).

\section{Analysis and Evaluation}

One way to demonstrate the impact of the temporal context in our social query expansion, is to compare a local strategy (computing contextual vectors for a specific time period, such as a single day) versus a global strategy (computing contextual vectors across all days) for a set of  hashtags. 

To compare the local versus global strategy for social query expansions, we look at the time range from 1-October-2016 to 3-November-2016 on Twitter (complete firehose access). According to either strategy, for every day each hashtag is expanded to a set of ngrams related to the hashtag, which are treated as queries. The hashtag and the ngram expansions are then used to find matching relevant document links.  For the purposes of this evaluation we are looking for matches in the links' titles and descriptions.

For every day in the time interval under consideration, we take a predetermined set of 20 hashtags and use their contextual vectors to get an expanded set of ngrams which are related to each hashtag. For each expanded hashtag, we find all the links from the same day whose titles or descriptions contain the hashtag, the word-broken hashtag, or any of the expanded ngrams. This process is repeated on a daily basis for both strategies. In the local strategy, we expand the hashtags of the day using the contextual vectors computed in that day. In the global strategy on the other hand, we expand the hashtags of the day using a fixed, global set of hashtag expansions, which is computed by merging all the daily hashtag expansion data and selecting the highest scoring expansions
for each hashtag across all days.

\begin{figure}[tp]
\centering
\includegraphics[scale=0.33]{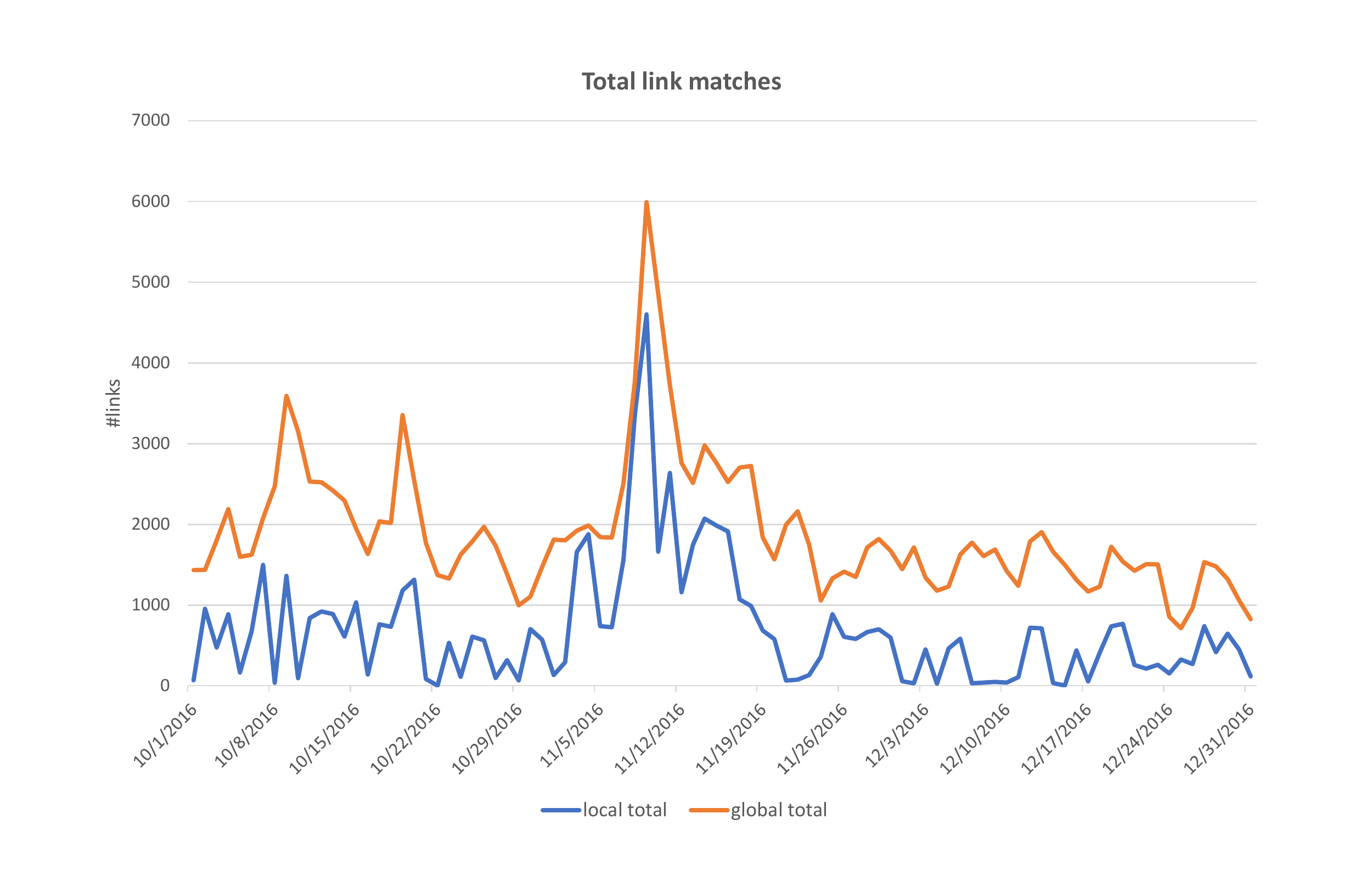}
\caption{ Total number of document link matches per day for the two strategies. }
\label{fig:all-results}
\end{figure}

\begin{figure}[tp]
\centering
\includegraphics[scale=0.33]{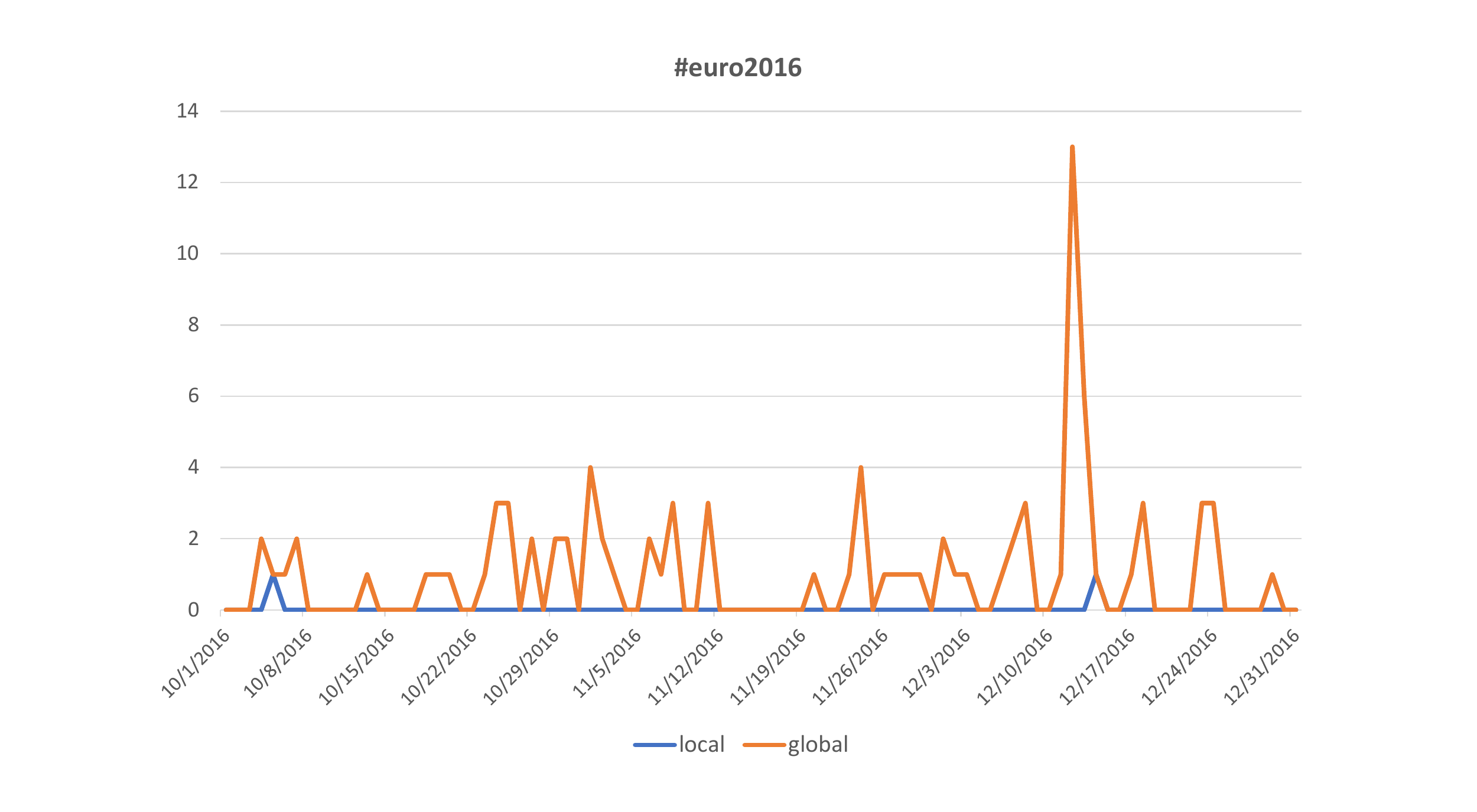}
\caption{Example where the global expansion produces a false positive peak for \texttt{\small {\#Euro2016}}, a few months after the event. 
The spike is caused by Ronaldo's golden boot award that is not part of the main tournament event. }
\label{fig:global-lose}
\end{figure}

We take the top 10 expansions of each day for every hashtag of that day for the local strategy and the top 10 expansions overall for the global strategy and count how many links are matched using these expansions. Figure~\ref{fig:all-results}  shows the total number of link matches per day for the two strategies. The global strategy finds more matched documents links than the local, so it appear that it performs better. However, the truth is that the global strategy, by using popular expansions every day, even when there is no related event in the news, finds many spurious matches that are not relevant to the hashtag. The recall is high, but the precision is low. This can be seen in Figure~\ref{fig:global-lose}, where the global strategy finds several matches for \texttt{\small{\#euro2016}} (the European football championship), but the local most of the time finds none. The global strategy expands the hashtag \texttt{\small{\#euro2016}} to ``Christiano Ronaldo'' (a famous football player) and then matches any links that happen to be about the football player, even if they have nothing to do with the Euro 2016 championship. The expansion to ``Christiano Ronaldo'' happened for only a single day, but the global strategy applies it to every day, so it has a high chance of matching links. The local strategy however applies the expansion only on that specific same day, so it does not find any spurious matches for the other days. 

\begin{figure}[tp]

\includegraphics[scale=0.38]{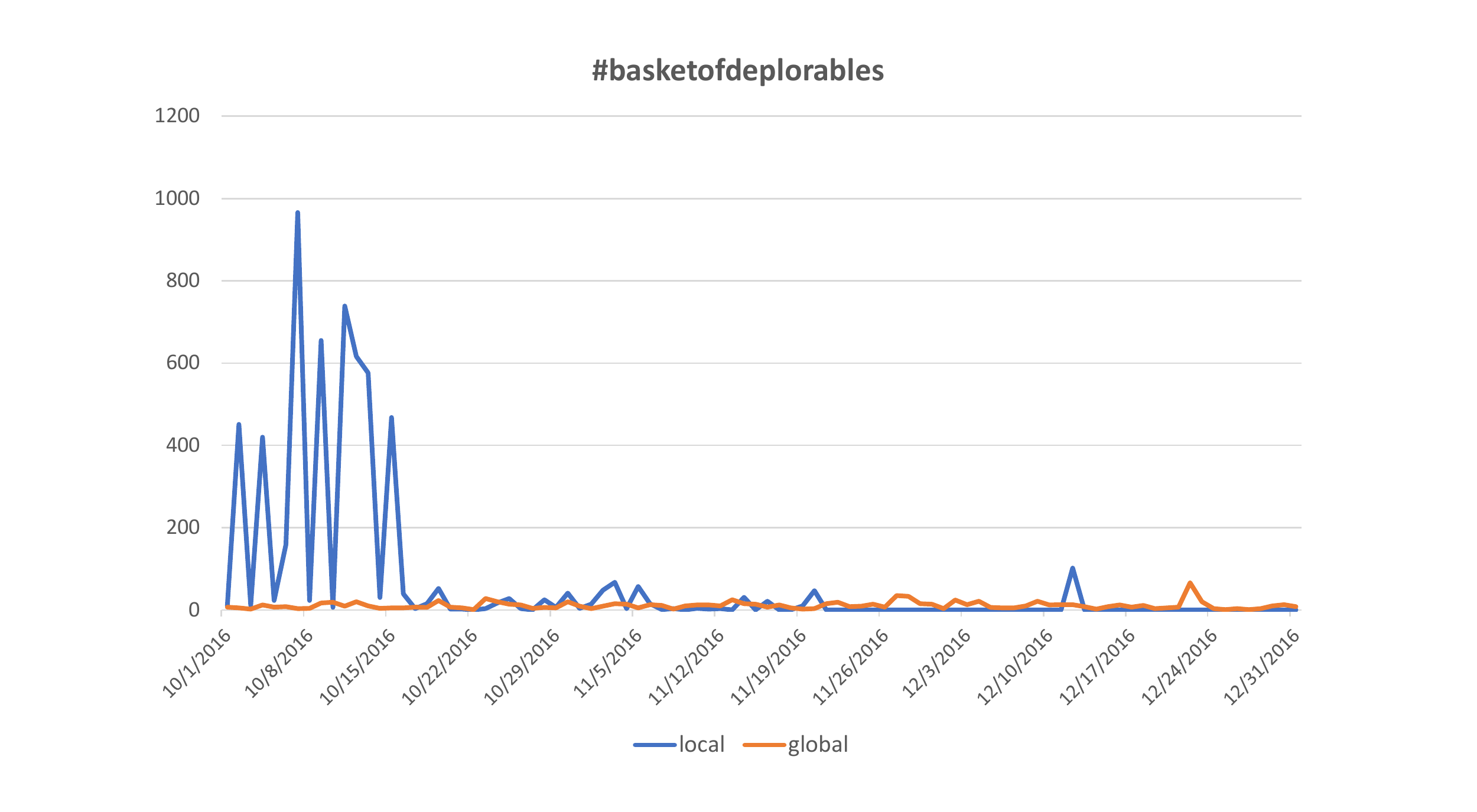}
\caption{Example where the local expansion technique provides a high number of matches compared to a global expansion for Clinton's famous phrase. }
\label{fig:global-wins}
\end{figure}

The fact that the local strategy uses the expansions computed in some day to find matches on the same day and not globally, helps it detect many articles that would otherwise be lost. This is because the articles could be about a specific aspect of the hashtag, which is frequently used on Twitter on some date, but a different aspect of the same hashtag becomes popular on a different date. The expansions in the local strategy can adapt and always capture the aspect that the Twitter users are referring to on each day. For example, \texttt{\small{\#basketofdeplorables}} (a term used by Hillary Clinton to refer to some supporters of Donald Trump) expands to ``Alicia Machado'' in one day and to ``David Duke'' in a different day, depending on which person was in the news at the time. The global strategy uses a fixed set of expansions (which do not include either of those two names) and therefore misses any articles about them. Figure~\ref{fig:global-wins} shows how the local strategy can significantly outperform the global one in this case. 
For the remaining days, when the hashtag was not actively used on Twitter, both strategies find only small numbers of matches. 
Computing different expansions every day helps us find the most relevant expansions based on what people are tweeting that day, which in turn helps us find more matching links, since the links are also more strongly associated with the events that are happening (and what people are discussing) that day.

\begin{figure}[tp]
\centering
\includegraphics[scale=0.35]{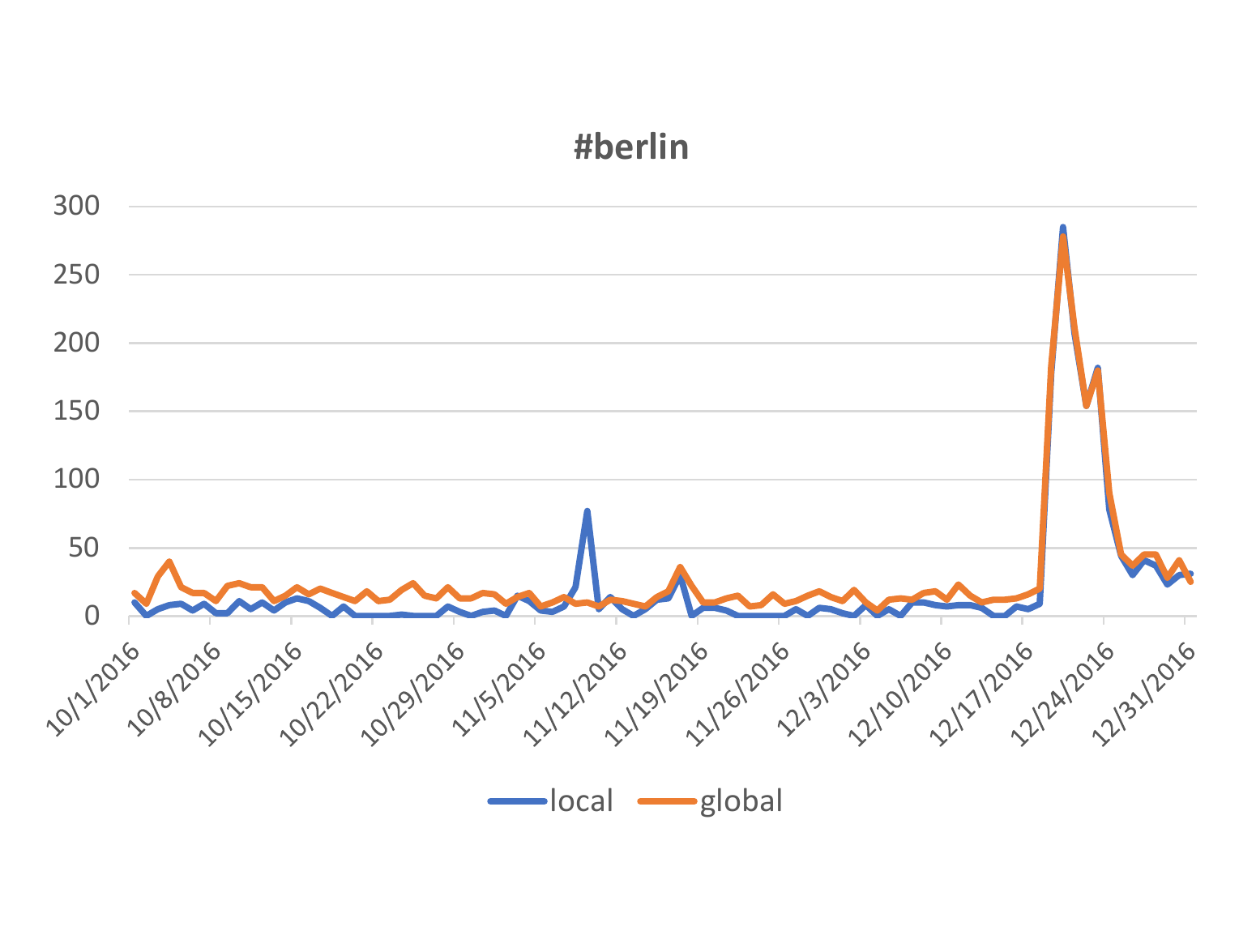}
\caption{Comparable amplitude values for local and global examples for Berlin attack in Germany.}
\label{fig:berlin}
\end{figure}

Another similar example is for \#berlin. The major event during the time range under consideration is the Berlin terrorist attack, which both strategies detect (Figure~\ref{fig:berlin}). This event overwhelms everything else and the global strategy cannot find many other matches. The local strategy however is able to find more matches for November 9 (the anniversary of the fall of the Berlin Wall), because it computes different expansions for that date. 

\begin{figure}[tp]
\centering
\includegraphics[scale=0.32]{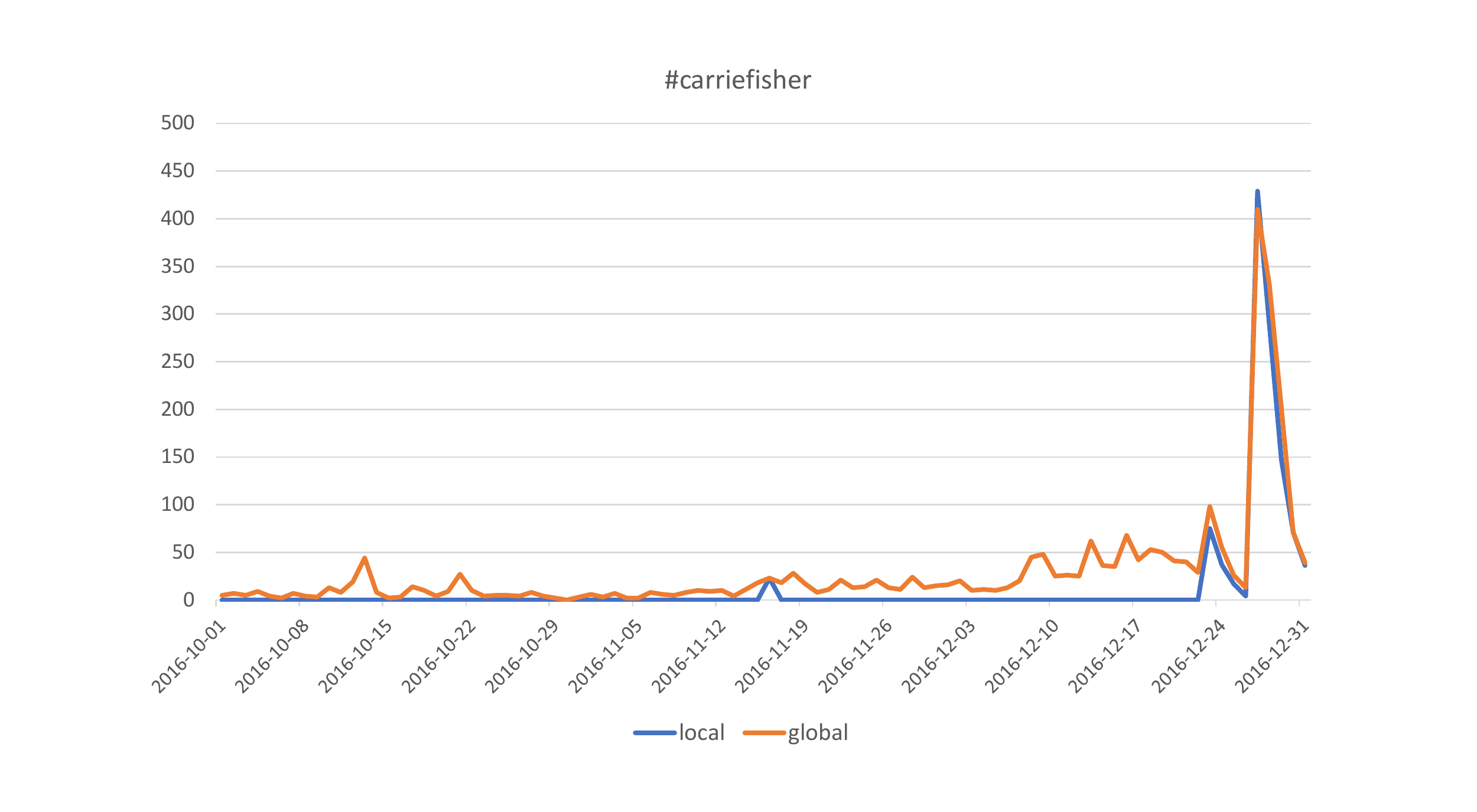}
\caption{Comparable amplitude values for local and global expansions for Carrie Fisher.}
\label{fig:equals}
\end{figure}

\begin{figure}[tp]
\centering
\includegraphics[scale=0.45]{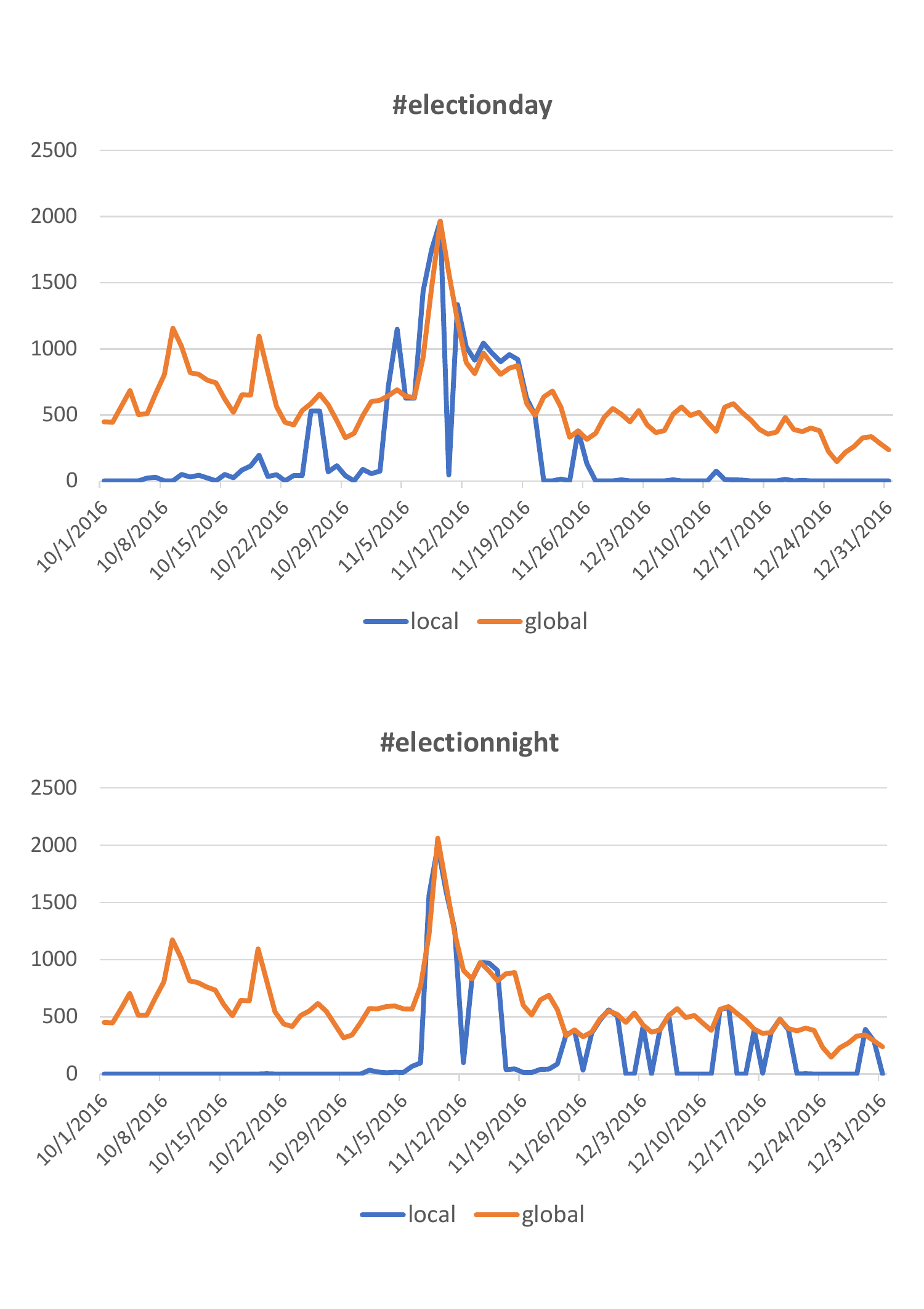}
\caption{A couple of examples for the main election event. 
 When local is low and global is high, global matches 
are off topic (the event has not happened yet), whereas when both are high the matches are relevant to the topic.}
\label{fig:elections}
\end{figure}

Sometimes both strategies detect a large number of articles. This happens when both strategies use similar expansions and so match mostly the same articles. The usual case for this is when an event happens once during the whole time range, like for example the death of Carrie Fisher
 (Figure \ref{fig:equals}). 
The actress was not mentioned during the time range we are considering other than because of her death. So, any expansions for \texttt{\small{\#carriefisher}} were computed from dates on or after her death and from tweets about her death. The set of expansions is therefore quite limited and unchanging, and thus both strategies end up using very similar expansions. Comparison of local vs. global for the US elections is
presented in Figure~\ref{fig:elections}.

We have identified the following possible behaviors of the two strategies:
\begin{enumerate}
\setlength{\parskip}{0pt}
\item The global strategy finds many matches, but the local does not (the example of \texttt{\small{\#euro2016}}, 
Figure~\ref{fig:global-lose}). Typically, the matches found by the global strategy are not relevant and should not be included as they would decrease precision.
\item The local strategy finds many matches, but the global strategy does not (the example of \texttt{\small{\#basketofdeplorables}}, Figure~\ref{fig:global-wins}). Typically, the local strategy has detected expansion terms that are popular that day and any matches it finds are useful in increasing the recall and should be included.
\item Both strategies find many matches
(e.g.,  \texttt{\small{\#carriefisher}}).

 This indicates that the hashtag is used for an uncommon event and any matches will be relevant and should be included.
\item Both strategies find a small number of matches. In this case, there is a high chance the matches are spurious and not relevant. 
Even if there are relevant matches, the signal is weak, which means not many people were using that hashtag at the time. 
Given the high risk of bad matches and the small benefit of a small number of relevant matches, we prefer not to include any matches.
\end{enumerate}

By monitoring how the two strategies work, we can decide whether to include or not any matching links and increase the recall of links without sacrificing precision.

\section{Conclusion}

In this paper we examined the use of shared links and hashtags for social query expansion.
We compared two strategies for social query expansion and found that the temporally sensitive social data performs better for query expansion than using a global set of expansions.
\balance
 
\bibliographystyle{ACM-Reference-Format}
\bibliography{sprf_analysis} 


\begin{thebibliography}{00}


\ifx \showCODEN    \undefined \def \showCODEN     #1{\unskip}     \fi
\ifx \showDOI      \undefined \def \showDOI       #1{#1}\fi
\ifx \showISBNx    \undefined \def \showISBNx     #1{\unskip}     \fi
\ifx \showISBNxiii \undefined \def \showISBNxiii  #1{\unskip}     \fi
\ifx \showISSN     \undefined \def \showISSN      #1{\unskip}     \fi
\ifx \showLCCN     \undefined \def \showLCCN      #1{\unskip}     \fi
\ifx \shownote     \undefined \def \shownote      #1{#1}          \fi
\ifx \showarticletitle \undefined \def \showarticletitle #1{#1}   \fi
\ifx \showURL      \undefined \def \showURL       {\relax}        \fi
\providecommand\bibfield[2]{#2}
\providecommand\bibinfo[2]{#2}
\providecommand\natexlab[1]{#1}
\providecommand\showeprint[2][]{arXiv:#2}

\bibitem[\protect\citeauthoryear{Alonso, Kandylas, and Tremblay}{Alonso
  et~al\mbox{.}}{2018}]%
        {AlonsoKT18}
\bibfield{author}{\bibinfo{person}{Omar Alonso}, \bibinfo{person}{Vasileios
  Kandylas}, {and} \bibinfo{person}{Serge{-}Eric Tremblay}.}
  \bibinfo{year}{2018}\natexlab{}.
\newblock \showarticletitle{{How it Happened: Discovering and Archiving the
  Evolution of a Story Using Social Signals}}. In \bibinfo{booktitle}{{\em
  Proceedings of {JCDL}}}. \bibinfo{pages}{193--202}.
\newblock


\bibitem[\protect\citeauthoryear{Dang and Croft}{Dang and Croft}{2010}]%
        {DangC10}
\bibfield{author}{\bibinfo{person}{Van Dang} {and} \bibinfo{person}{W.~Bruce
  Croft}.} \bibinfo{year}{2010}\natexlab{}.
\newblock \showarticletitle{Query Reformulation Using Anchor Text}. In
  \bibinfo{booktitle}{{\em Proc. of WSDM}}. \bibinfo{pages}{41--50}.
\newblock


\bibitem[\protect\citeauthoryear{Miyanishi, Seki, and Uehara}{Miyanishi
  et~al\mbox{.}}{2013}]%
        {MiyanishiSU13}
\bibfield{author}{\bibinfo{person}{Taiki Miyanishi}, \bibinfo{person}{Kazuhiro
  Seki}, {and} \bibinfo{person}{Kuniaki Uehara}.}
  \bibinfo{year}{2013}\natexlab{}.
\newblock \showarticletitle{Improving Pseudo-relevance Feedback via Tweet
  Selection}. In \bibinfo{booktitle}{{\em Proc. of CIKM}}.
  \bibinfo{pages}{439--448}.
\newblock


\bibitem[\protect\citeauthoryear{Wang, Huang, and Feng}{Wang
  et~al\mbox{.}}{2017}]%
        {WangHF17}
\bibfield{author}{\bibinfo{person}{Yashen Wang}, \bibinfo{person}{Heyan Huang},
  {and} \bibinfo{person}{Chong Feng}.} \bibinfo{year}{2017}\natexlab{}.
\newblock \showarticletitle{Query Expansion Based on a Feedback Concept Model
  for Microblog Retrieval}. In \bibinfo{booktitle}{{\em Proc. of {WWW}}}.
  \bibinfo{pages}{559--568}.
\newblock


\bibitem[\protect\citeauthoryear{Xu and Croft}{Xu and Croft}{2000}]%
        {Xu00}
\bibfield{author}{\bibinfo{person}{Jinxi Xu} {and} \bibinfo{person}{W.~Bruce
  Croft}.} \bibinfo{year}{2000}\natexlab{}.
\newblock \showarticletitle{Improving the Effectiveness of Information
  Retrieval with Local Context Analysis}.
\newblock \bibinfo{journal}{{\em {ACM} Trans. Inf. Syst.\/}}
  \bibinfo{volume}{18}, \bibinfo{number}{1} (\bibinfo{year}{2000}),
  \bibinfo{pages}{79--112}.
\newblock


\bibitem[\protect\citeauthoryear{Zingla, Latiri, Mulhem, Berrut, and
  Slimani}{Zingla et~al\mbox{.}}{2018}]%
        {ZinglaLMBS18}
\bibfield{author}{\bibinfo{person}{Meriem~Amina Zingla},
  \bibinfo{person}{Chiraz Latiri}, \bibinfo{person}{Philippe Mulhem},
  \bibinfo{person}{Catherine Berrut}, {and} \bibinfo{person}{Yahya Slimani}.}
  \bibinfo{year}{2018}\natexlab{}.
\newblock \showarticletitle{Hybrid query expansion model for text and microblog
  information retrieval}.
\newblock \bibinfo{journal}{{\em Inf. Retr. Journal\/}} \bibinfo{volume}{21},
  \bibinfo{number}{4} (\bibinfo{year}{2018}), \bibinfo{pages}{337--367}.
\newblock


\end{thebibliography}

\end{document}